\documentclass[graybox]{svmult}

\usepackage{mathptmx}       
\usepackage{helvet}         
\usepackage{courier}        
\usepackage{type1cm}        
\usepackage{makeidx}         
\usepackage{graphicx}        
\usepackage{multicol}        
\usepackage[bottom]{footmisc}

\makeindex             

\begin{document}

\title*{Population Synthesis of Massive Close Binary Evolution}
\author{J.J. Eldridge}
\institute{J.J. Eldridge \at The Department of Physics, The University of Auckland, Private Bag 92019, Auckland, New Zealand, \email{j.eldridge@auckland.ac.nz}}
%
%
\maketitle

\abstract*{Binary population synthesis is the method by which
  predictions of varied observables of stellar populations can be made
  from theoretical models of binary stellar evolution. Binary stars
  have many more possible evolutionary outcomes compared to single
  stars and the relative rates of the different pathways, such as the
  rates of different supernova types, depend on uncertain or poorly
  constrained physics.  In this Chapter we describe population
  synthesis, outline the major uncertainties and discuss the relevant
  predictions for core-collapse supernovae.  After we overview single
  star evolution we outline the important physical processes that
  occur in binaries including Roche-lobe overflow, common-envelope
  evolution and supernova kicks. We also discuss how a synthetic
  stellar population incorporating interacting binaries can be
  constructed and how uncertainties, such as the strength of supernova
  kicks, affect any predictions. We illustrate the process by
  comparing predictions for the stellar populations in two young star
  clusters. We then discuss the important predictions from population
  synthesis for understanding core-collapse supernovae, their
  delay-time distribution and their progenitor stars. Finally we
  discuss how we can predict the rate of mergers of compact remnants
  and thus predict the initial parameters of gravitational wave
  sources.}

\abstract{Binary population synthesis is the method by which
  predictions of varied observables of stellar populations can be made
  from theoretical models of binary stellar evolution. Binary stars
  have many more possible evolutionary outcomes compared to single
  stars and the relative rates of the different pathways, such as the
  rates of different supernova types, depend on uncertain or poorly
  constrained physics.  In this Chapter we describe population
  synthesis, outline the major uncertainties and discuss the relevant
  predictions for core-collapse supernovae.  After we overview
  single star evolution we outline the important physical processes
  that occur in binaries including Roche-lobe overflow,
  common-envelope evolution and supernova kicks. We also discuss how a
  synthetic stellar population incorporating interacting binaries can
  be constructed and how uncertainties, such as the strength of
  supernova kicks, affect any predictions. We illustrate the process
  by comparing predictions for the stellar populations in two young
  star clusters. We then discuss the important predictions from
  population synthesis for understanding core-collapse supernovae,
  their delay-time distribution and their progenitor stars. Finally we
  discuss how we can predict the rate of mergers of compact remnants
  and thus predict the initial parameters of gravitational wave
  sources.}

\section{Introduction}
\label{sec:1}

Supernovae arise from stars. Therefore to understand these explosive
death throes we must understand the evolution of stars. While studying
a single supernova and its associated progenitor star can be useful,
only a handful of events have suitable observational datasets to
understand them. An alternative method to gain understanding of
supernovae is to study the population of explosions. To do this the
population of supernovae must be modelled. This requires creating a
synthetic stellar population of the stellar progenitors that can be
compared to the observed population of supernovae. This is performed
by population synthesis.

One of the main problems of population synthesis is how to ensure the
accuracy of an individual stellar model and the resultant
population. This is especially true when attempting to model
supernovae as every phase of evolution must be calculated and thus
many different uncertainties can limit the confidence of any
predictions.  One direct method is to compare models to the Sun or
other well observed stars. For the latter case the best examples are
double-lined spectroscopic-eclipsing binary stars, because for these
stars it is possible to accurately deduce their mass, radius and
surface temperature of the stars \cite{torres,southworth}. However
these primarily only provide tests of stellar structure, with many
containing only main-sequence stars, and are thus limited in what
they can tell us about the broader picture of stellar evolution.

To achieve a broader view the next logical step is to study
populations of stars, such as those in a star cluster with similar
ages and initial compositions. The different evolutionary states of
stars in a population is due to their initial mass. Population
synthesis is the method of combining stellar models of different
initial masses (and other initial parameters such as composition) to
make a prediction of how an entire stellar population would appear at
various ages. This would be similar to a star cluster. In contrast a
mixture of stellar populations with a range of ages would simulate the
stellar population of a galaxy.

The general result we find from population synthesis is that most
stars are found on the main-sequence when they fuse hydrogen to
helium. We then expect stars to evolve into red supergiants, white
dwarfs and the other stellar types depending on their initial mass,
composition and mass loss. We find that the luminosity of the
brightest stars on the main sequence depends on the age of the cluster
as the most luminous stars are the first to evolve off the
main-sequence.

This simple picture is complicated by binaries. Binary interactions
can prevent certain evolutionary pathways; for example, preventing
stars from becoming red giants by removing their hydrogen
envelopes. However this opens up alternative pathways not available to
single stars. Recent work has revealed the importance of binary
evolution in accurately modelling stars and stellar populations
\cite{demarco}. Rather than becoming red giants, the exposed cores
evolve to become helium-rich dwarfs with significantly higher surface
temperatures. When one star loses its hydrogen envelope the companion
star can accrete the material, increasing the mass of the star. In
extreme cases the two stars can merge. These processes can upset the
simple relationship between the most massive and luminous stars and
the age of a stellar population.

Adding to this complexity, new evolutionary pathways introduce greater
variation due to an increased number of initial parameters. In binary
systems we must account for the mass of both stars and their initial
separation. These will sensitively determine when the stars will
interact and the severity of this interaction. Finally the physics of
binary evolution is more uncertain than single star evolution because
it is inherently more complex and based on short-timescale dynamical
processes.

In this chapter we outline these aspects of population synthesis in
more detail and pay special attention to uncertainties and
complexities that are still being uncovered. We then discuss some
recent and interesting observational constraints concerning
supernova. These include their relative rates, their delay-time
distribution, observed supernova progenitors and the detection of
merging compact remnants.

\section{Stellar evolution}

Even though the Sun and other stars appear to be unchanging on human
timescales the process of nuclear fusion at their centre is slowly
using up their energy supply. Over millions and billions of years
stars continue to change and evolve until they use up all their
nuclear fuel. Before describing how a population of stars is expected
to evolve, we must first review the basics of how stars evolve.

\subsection{Single stars}

Stars are spherical fluid objects in hydrostatic equilibrium. Their
cores are hot enough to undergo nuclear fusion. All stars initially
burn hydrogen to helium and then evolve off the main sequence and burn
successively heavier elements in their cores; the more massive the
star the heavier the elements it is able to fuse. Low-mass stars form
only a helium core. Intermediate-mass stars can fuse helium to form a
carbon-oxygen core. Massive stars can fuse up to a core of iron-group
elements.

For low and intermediate mass stars, once these cores are formed their
evolution is normally over; their hydrogen envelopes will be lost and
those cores go on to become white dwarfs. The massive stars that have
formed iron cores will explode in a core-collapse supernova typically
leaving a neutron star or black hole remnant.

While burning hydrogen, stars are normally small and compact
main-sequence stars. They grow in radius to become red giants once a
helium core is formed. One aspect that greatly affects the evolution
of stars is mass loss in stellar winds. Only massive stars, or other
stars near the end of their lives, have strong stellar winds that can
effect the evolution of a star. The evolutionary pathway taken for
single stars is a competition between the progression of nuclear
burning at the centre of the star and mass loss from the surface.

For low and intermediate mass stars, mass loss becomes strong when the
star becomes a large red giant. These stars have a helium or
carbon-oxygen core and the envelope loss results in a white dwarf. For
massive stars, mass loss has the key impact on determining the nature
of their observed supernova. For example if mass loss is weak then the
supernova will be hydrogen-rich (type II). When the winds are strong
enough to remove the hydrogen and leave only the helium core, the
supernova would be hydrogen-poor (type Ib/c). We summarise the
expected evolutionary pathways and their corresponding mass ranges at
Solar metallicity in Table \ref{evolution}.

The values in Table \ref{evolution} vary for stars of different
initial metallicities or rotation rates. Changes are more significant
for massive stars that will go to core-collapse. If a star is rotating
fast enough to induce a strong mixing effect it forms a larger core
and evolves as a slightly more massive star. This effect will be
biggest in reducing the initial mass for type Ib/c
supernovae. Typically these differences are only a few
$M_{\odot}$. The differences can be much larger when metallicity is
varied. For example, if the metallicity is decreased this leads to
weaker stellar winds that have the opposite effect and so will
increase the minimum mass for a type Ib/c supernova from a
non-rotating star to an initial mass of 80$M_{\odot}$ at 1/10th Solar
metallicity.

\begin{figure}[t]
\sidecaption[t]
\includegraphics[scale=.5]{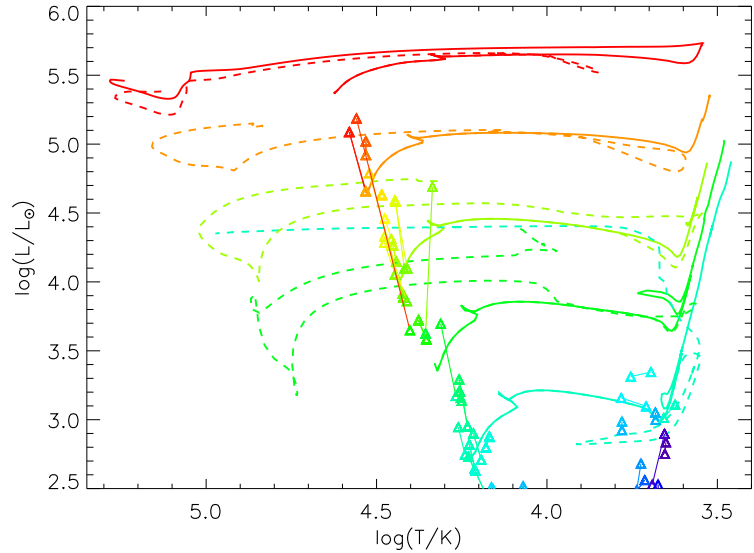}
\caption{An example Hertzsprung-Russell diagram showing massive star
  evolution tracks and observed eclipsing binary stars for which mass,
  temperature and luminosity can be accurately determined
  \cite{southworth}. The shading represents the mass of the stars,
  solid line tracks are for single stars and dashed tracks are for
  binary stars. The masses of the tracks are 5, 7.5, 12, 20 and
  40M$_{\odot}$.}
\label{fig1}     
\end{figure}

We show some sample evolution tracks on a Hertzsprung-Russell diagram
in Figure \ref{fig1}, in which a star's luminosity and surface
temperature are related. On Figure \ref{fig1} we include stellar
models and observed eclipsing binary stars (where the parameters are
well known) for comparison. We can see many of the stars are on the
main sequence. Relatively few post-main sequence binaries are known
because the post-main sequence lifetime is short and the stars in
binaries are more likely to have interacted. This simple diagram
demonstrates that we can have confidence in stellar models as they
reproduce at the same time the observed masses, luminosities and
surface temperatures.

\begin{table}[!b]
\caption{Mass ranges for typical evolution of single stars, along with
  nuclear burning reactions during each of the evolutionary
  states. The different phases are the main sequence (MS) when stars
  are hydrogen burning the other types occur post-main sequence: red
  giant branch (RGB), asymptotic giant branch (AGB), super-AGB (SAGB),
  red supergiants (RSG), Wolf-Rayet (WR) stars, blue supergiants (BSG)
  and luminous-blue variables (LBV). We also include two types of
  supernovae (SNe) here, type II with hydrogen and type Ib/c without
  hydrogen in the spectra. The remnants are either a helium white
  dwarf (HeWD), a carbon-oxygen white dwarf (COWD), a oxygen-neon
  white dwarf (ONeWD), a neutron star (NS) or a black holes (BH). The
  values are for Solar metallicity and the values due vary with
  metallicity, initial rotation rate and binary interactions as
  discussed in the text.}
\label{evolution}      
\begin{tabular}{p{1.5cm}p{7.8cm}p{2cm}}
\hline\noalign{\smallskip}
$M_{\rm initial}/M_{\odot}$ & Evolution pathway & Total Lifetime   \\
\noalign{\smallskip}\svhline\noalign{\smallskip}
$<$0.08 & Mass at centre of star too low for nuclear fusion, the object is a brown dwarf rather than a star. & -- \\
0.08--0.8 & MS $\rightarrow$ RGB $\rightarrow$ HeWD  & 4500--29 Gyrs \\
      & H-burning only $\rightarrow$ Strong stellar-wind removes envelope.\\
0.8--6.5 & MS $\rightarrow$ RGB $\rightarrow$ AGB $\rightarrow$ COWD  & 29Gyrs--66Myrs \\
      & H-burning $\rightarrow$ He-burning $\rightarrow$ Strong stellar-wind removes envelope.\\
7--8 & MS $\rightarrow$ RGB $\rightarrow$ AGB $\rightarrow$ SAGB $\rightarrow$ ONeWD  & 66--41 Myrs \\
      & H-burning $\rightarrow$ He-burning $\rightarrow$ C-burning $\rightarrow$ Strong stellar-wind removes envelope.\\
8--20 & MS $\rightarrow$ RSG $\rightarrow$ Type-II SN $\rightarrow$ NS  & 41--9.7 Myrs \\
      & H-burning $\rightarrow$ Fusion progresses to an iron-group elements core.\\
20--25 & MS $\rightarrow$ RSG $\rightarrow$ Type-II SN $\rightarrow$ BH  & 9.7--7.7 Myrs \\
      & H-burning $\rightarrow$ Fusion progresses to an iron-group elements core.\\
25--40 & MS $\rightarrow$ RSG $\rightarrow$ WR $\rightarrow$ Type-Ib/c SN $\rightarrow$ BH  & 7.7--5.1 Myrs \\ 
      & H-burning $\rightarrow$ Fusion progresses to an iron-group elements core $\rightarrow$ Strong stellar-wind removes envelope.\\
40--100 & MS $\rightarrow$ BSG/LBV $\rightarrow$ WR $\rightarrow$ Type-Ib/c SN $\rightarrow$ BH  & 5.1--3.1 Myrs \\ 
      & H-burning $\rightarrow$ Fusion progresses to an iron-group elements core $\rightarrow$ Strong stellar-wind removes envelope..\\
\noalign{\smallskip}\hline\noalign{\smallskip}
\end{tabular}
\end{table}

\subsection{Binary stars}

A binary star system is one in which two stars are bound together by
their mutual gravity, orbiting around their common centre of mass. The
mass, period and orbital separation are related by Kepler's Third Law
which can be expressed as,
\begin{equation}
\frac{(M_1 + M_2)}{M_{\odot}} = \frac{(a / {\rm A.U.})^3}{(P / {\rm yr})^2} = \frac{(a / {\rm 215\, R_{\odot}})^3}{(P / {\rm 365.25 \, days})^2};
\end{equation}
where $M_1$ and $M_2$ are the masses of the two stars in the binary,
$P$ is the orbital period in years or days and $a$ is the orbital
separation in astronomical units or Solar radii.

For a star in a binary there is the possibility of extra physical
processes. For example, tidal forces exchanging angular momentum
between the stars' rotation and their orbit. This effect may be
enhanced if magnetic fields are present in the stars. However the most
important extra process is the possibility of mass transfer that leads
to both enhanced mass loss and mass gain for the stars in a binary.

Using the mass ranges in Table \ref{evolution} as a guide, enhanced
mass loss in binary systems means that any star initially more massive
than 10$M_{\odot}$ can now lose its hydrogen envelope and give rise to
a type Ib/c supernova. While mass gain by stars means that, for
example in an extreme case, a binary with two 10$M_{\odot}$ stars
could merge and form a 20$M_{\odot}$ star. This would happen at ages
up to 25Myrs, the main-sequence lifetime of the 10$M_{\odot}$
stars. Therefore the 20$M_{\odot}$ might appear to be surrounded by an
older population than its typical 9.7Myrs lifetime. Many clusters have
such stars in the population as blue stragglers.

In the simplest and extreme view, one can expect that if a star's
radius grows to be of the order of the orbital radius the stars will
``get in each other's way''. In reality the interaction happens before
a star grows to the size of the orbit. In the rotating frame of the
binary, along the line between the centre of the two stars there is a
point where the gravity of the two stars is equal. If stellar material
of one star goes beyond this point, the other star's gravity will
dominate and pull the material towards it. The surface where the
gravity of the two stars cancels is called the Roche Lobe. Nearly
every stellar evolution code assumes spherical symmetry, so
approximate expressions are calculated to estimate the radius, $R_{\rm
  L}$, at which a star fills its Roche Lobe. The most widely used one
was derived by Eggleton \cite{eggleton},
\begin{equation}
R_{\rm L,1}/R_{\odot} = \frac{ 0.49 q^{2/3}}{0.6 q^{2/3} + \ln (1+q^{1/3})},
\end{equation}
where $q=M_1/M_2$ is the mass ratio of the two stars. If a star fills
this volume it can be assumed that any mass beyond this volume flows
from the star to its companion. This is referred to as Roche-Lobe
Overflow (RLOF) and leads to both mass loss and mass gain for the
primary and secondary stars respectively. This mass exchange can be
considerably stronger than the typical mass loss driven by stellar
winds.

During this mass transfer angular momentum is also transferred. This
can lead to the companion star being spun up to almost critical
rotation rates\cite{demink1}. In fact the most rapidly rotating stars
are likely to be the result of binary evolution. Those stars that are
spun up can experience evolutionary pathways that could lead to
long-GRBs if they also become massive enough\cite{cantiello}.

If RLOF does not prevent the radius of the overflowing star from
growing it may eventually engulf the other, this leads to
common-envelope evolution (CEE). Here the core of one star and its
companion now orbit around one another within the envelope of the
larger star. The exact details of such evolution are extremely
uncertain. We know it must occur because of the large number of white
dwarfs in very close binary systems with periods of days. These are only
possible if the system's excess angular momentum has been carried away
by an ejected common envelope. Our theoretical understanding of CEE is
still at its beginning\cite{ceereview}. We do understand that there
are two possible outcomes from CEE, either for the two stars to merge
or for the formation of a much tighter binary, with one of the two
stars having lost its hydrogen envelope.

In comparison to CEE, RLOF is relatively well understood as there
are a number of observed systems where this process is occurring. One
of the better known systems for example is $\beta$-Lyrae where stable
mass-transfer is still happening today at the rate of approximately
$2\times10^{-5}\, M_{\odot}\,{\rm yr^{-1}}$\cite{betalyrae}.

The key issues of mass transfer in population synthesis concern
whether a binary systems only experiences RLOF or whether the
interaction progresses to CEE. CEE can only be avoided if mass loss
can prevent the star from expanding and also if mass transfer causes
the orbit to shrink or grow. Typically if the more massive star is the
donor then the orbit shrinks on mass transfer which can lead to
CEE. However if the transfer is not fully conservative and mass is lost
the system can be prevented from shrinking and
entering CEE. If the less massive star is the donor the orbit widens.

The evolutionary state of a star can also affect mass transfer
stability. Typically mass transfer during the main-sequence is driven
at a nuclear evolutionary timescale so mass transfer is slow and
stable. For post-main sequence mass transfer the evolutionary
timescale is shorter so CEE becomes more likely. We note that tidal
forces at the time of interactions can also be important. As one star
grows it can be spun up at the expense of angular momentum from the
orbit, again possibly instigating CEE. Many of these factors are
relatively well understood. The one that is uncertain and most
important is how efficient or conservative mass-transfer is. It
appears that while, in general, mass-transfer is efficient between
stars with similar initial masses there are more complex factors at
work that we still need to understand
\cite{masstransfer1,masstransfer2}.

While we have limited understanding of CEE it is typically
parametrised in population synthesis\cite{ceereview,hurley} by
comparing the binding energy of the overflowing star's envelope to the
orbital energy of the core of the primary star and its companion,
\begin{equation}
E_{\rm bind,i} = \alpha_{\rm CE} (E_{\rm orb,f} - E_{\rm orb,i}).
\end{equation}
Where,
\begin{equation}
E_{\rm bind,i}= \frac{-GM_1 M_{\rm env, 1}}{ \lambda R_1},
\end{equation}
is the initial binding energy of the envelope of the star with
$\lambda$ being a constant depending on the structure of the star and
$M_{\rm env, 1}$ being the mass in the envelope of the donor
star. Then,
\begin{equation}
E_{\rm orb}= \frac{-GM_{c,1} M_2}{2 a},
\end{equation}
is the initial or final orbital energy of the primary's core of mass
$M_{\rm c,1}$ and the secondary star. The free-parameter in the model
is $\alpha_{\rm CE}$ and is a constant of the order 1. While there are
subtle variations on this model they all effectively use conservation
of energy. Although a model including angular momentum of the stars
and orbit has also been considered\cite{angmvcee}. Of course if there
is a significant amount of binding energy then CEE can lead to the
merging of the two stars.

In Figure \ref{fig1} we show examples of evolutionary tracks of
stars that experience RLOF or CEE compared to the single star
tracks. The main result is that the stars lose their hydrogen envelope
and undergo further evolution as helium stars or Wolf-Rayet
stars. The greatest difference compared to the single-star
evolution in Figure \ref{fig1} and Table \ref{evolution} is for stars
between 8 to 25$M_{\odot}$. For these stars stellar winds are not
strong enough to remove the hydrogen envelope at any stage of
evolution. Therefore only RLOF or CEE can remove the hydrogen envelope
to create helium stars. Stars in the other mass ranges might have
their mass loss accelerated but the eventual outcome is not strongly
affected.

Typically in a binary system the initially more massive star evolves
faster and hence is the first to interact. Whether the lower mass companion
also experiences RLOF or CEE depends on whether the binary is
disrupted or not. A binary can be disrupted if, for circular orbits,
half of the total mass of the binary is ejected from the system over a
short timescale. For low and intermediate-mass stars this can be
unlikely. But in massive stars a supernova can rapidly eject a large
amount of mass and can unbind the binary.

The loss of half the mass in a binary leads to the end of the binary,
as the system does not have enough mass for gravity to hold it
together. The compact remnant and the companion star will therefore
fly apart and continue their evolution as single stars.  The extreme
cases in this scenario are for type Ia or pair-instability supernovae
where the exploding star is disrupted. In these cases one star is
totally destroyed, leaving the other alone.

The complicating factor in this picture is that there is strong
observational evidence that, in supernovae, neutron stars and black
holes receive kicks when they are born \cite{nskicks,bhkicks}. The
exact reason for the kick is likely due to asymmetries during
core-collapse and the subsequent explosion. We know that some neutron
stars can have velocities up to 1000~$\rm km \,s^{-1}$, although the
distribution is similar to that of a Maxwell-Boltzmann distribution
with a characteristic velocity of 265~$\rm km \, s^{-1}$. For black
holes there is less direct evidence of how strong the kicks should be
but it is probable that their kicks are weaker \cite{bhkickslow}.

These supernova kicks can either unbind a binary that would have
remained bound or ensure a binary remains bound rather than
unbound. Of course this depends sensitively on the strength of the
kick and the direction.  When the binary remains bound further
evolution can lead to binary interactions between the secondary star
and the primary star's compact remnant. These systems are observed as
X-ray binaries. To first order the physics of these systems is similar
to that of binaries with two normal stars. The two important
differences are that the compact remnants tend to accrete smaller
amounts of material and are luminous in X-rays. These lead to more
stable mass transfer and the irradience of the companion star
affecting mass transfer respectively. The main uncertainty is how much
mass can be accreted because that determines the efficiency of mass
transfer and whether CEE happens in these binaries or not. In such
systems a second supernova can also occur giving rise to either two
runaway compact remnants or a double compact remnant binary that may
be observed as gravitational wave sources if they ultimately merge.

Because the magnitude and direction of kicks are usually assumed to be
random with respect to the position of the exploding star, multiple
kicks must be simulated with each binary model to predict the full
possible range of possible outcomes from a supernova. This is one of
the reasons why population synthesis was created -- the evolution of a
binary is no longer linear and single valued. An element of randomness
enters when a supernova occurs. This is again another reason why so
many computations are required in binary population synthesis as all
the possible evolutionary avenues must be explored and accounted for
statistically.

\section{Making a synthetic stellar population}

While individual stellar models are useful and can be used to
constrain our understanding of stellar evolution the next step is to
combine these models together to predict the appearance of a
population of stars. First we discuss how to do this with single star
models and then explain how it becomes more complex with a binary
population.

\subsection{Single stars}

\begin{figure}[t]
\sidecaption[t]
\includegraphics[scale=.5]{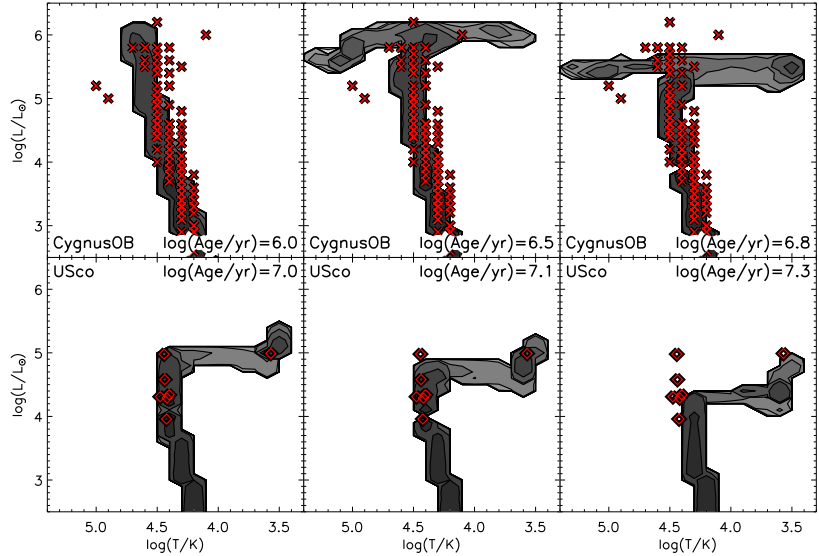}
\caption{Example Hertzsprung-Russell diagrams showing a synthetic
  single-star population compared to the observed populations of two
  stars clusters, the Cygnus OB association\cite{wright} in the upper
  panels in crosses and USco association \cite{pecaut} in the low
  panels in diamonds. The age of the population can be estimated by
  comparing the distribution of the most luminous stars. Each contour
  is separated by an order of magnitude in the probability density of
  stars on the HR diagram.}
\label{fig3}     
\end{figure}

Using single-star models, population synthesis is relatively
straightforward. We take stellar evolution tracks as shown in Figure
\ref{fig1}, which demonstrate how a star of a single mass evolves over
time, and combine them in proportions given by an estimated
distribution of initial masses to indicate how stars of different
masses look at the same age. We show such a combination for single
stars in Figure \ref{fig3}. In the Figure we illustrate how synthetic
populations appear at different ages and compare them to observed
clusters. The contours represent the probability of stars being
observed in a certain region of the Hertzsprung-Russell diagram. We
can see that as the population ages we eventually find red
(super)giants appearing and the main sequence shortens, with the most
massive stars evolving faster than the less massive stars. Matching
models to the distribution of turn-off and post main-sequence stars in
individual clusters, we find that the age of Cygnus OB is closest to
3Myrs and Upper Scorpius 10Myrs. We note that normally for single
stars isochrones are created which are lines of constant age but
varying mass. Here we have calculated density/contour plots so we can
later compare them to binary models where a simple isochrone is not
possible.

To create such plots from the stellar models, the appearance of all
possible stellar models at a specific age are recorded and any gaps in
the model mass distribution are interpolated over. This is less simple
than it sounds. One problem is the vast range of rates of evolution
for stars of different masses. For example a 100$M_{\odot}$ star will
evolve over 3Myrs compared to a 1$M_{\odot}$ star's lifetime of
10Gyrs. The post-main sequence lifetime of such stars might be only
10\% of their total lifetime which makes it difficult to account for
the shortest phases of evolution. Therefore either care must be taken
in interpolation between the different stellar models or some binning
of the models must be used as in Figure \ref{fig3}. Here we use time
bins of width 0.1dex in the logarithmic age of the stellar population.

The second problem we must deal with is that we must account for 
how many stars there are of each mass. This is typically referred to
as the initial-mass function (IMF) which describes how many stars
there are of different masses. The most basic form was first derived
by Salpeter \cite{salpeter} who proposed a simple power-law,
\begin{equation}
\frac{dN}{dM} \propto M^{-2.35},
\end{equation}
where $N$ is the number of stars of mass $M$. Much work has been
performed over the last few decades to determine the IMF. The most
commonly used one now is by Kroupa \cite{kroupa}. It is similar to
Salpeter at high mass but becomes shallower below $0.5M_{\odot}$ to
such that $\frac{dN}{dM} \propto M^{-1.35}$ down to a minimum mass of
$0.1M_{\odot}$. The IMF does also continue down to lower masses but
this is the domain of brown dwarfs and planets.  There are also other
forms \cite{imfreview}, but all tend to have the same high mass slope
but vary at the lower masses.

Both the IMF and the longer evolutionary times imply that there are
many more low-mass stars than high-mass stars. This compounds the
problems with interpolation as the fainter coolest stars are the most
numerous but the rarest, brightest stars dominate the appearance of a
cluster when it is young. There are solutions to these problems. The
interpolation and careful selection of the initial masses of the
stellar models used to create the populations are key.

For a simple demonstration of how important this is we can combine the
above IMF with an approximation of how the lifetime of a star depends
on mass. From homology we can prove that for most main-sequence stars
with radiative interiors, supported by gas pressure, $L\propto M^3$
and so the main sequence lifetime is given by, $\tau \propto
M^{-2}$. Therefore incorporating this equation with the IMF we find
that in a stellar populations that continuously forms stars the number
of stars at a given mass will be $\propto M^{-4.35}$. However given
that the more massive stars are more luminous, the contribution of
stars to the luminous output of a galaxy is not quite so skewed to the
lowest mass stars. The luminosity contribution of a star is $\propto
M^{-1.35}$ assuming continuous star formation and $\propto M^{0.65}$
for a very young star cluster.

For a single star population other parameters that
can be varied are initial composition, rotation rate and the scheme of
mass-loss rates employed when the stellar models are calculated. For
example the tracks in Figure \ref{fig1} and \ref{fig3} are assumed to
be Solar metallicity which is taken to be $Z=0.02$ (although there is
some debate on exactly what Solar metallicity is
\cite{solar1}). This just determines which generation of
stars the models represent. If $Z=0$ they would be the
first generation of stars in the Universe. The amount of hydrogen in
the models would also vary from $X=0.70$ to 0.75 respectively. The
remaining amount of mass in the stars will be made up of helium.

Changing the metallicity has strong effects on the opacity of the
stellar material and thus the stellar radii. Crucially, it also
affects the mass-loss rates in stellar winds. Most winds are driven by
radiation pressure from the iron lines in the atmosphere. Therefore
higher metallicities have more iron which subsequently leads to
stronger stellar winds.  Stellar winds become strongest for the most
luminous stars and generally have the dramatic effect of removing a
star's hydrogen envelope. For low- and intermediate-mass stars this
leads to RGB and AGB stars which later become white dwarfs. For the
massive stars it leads to hydrogen-poor Wolf-Rayet stars.

Unfortunately the mass-loss rates employed in a stellar model are
mostly empirical with only a few based on theoretical
calculations. Hence some changes can be made in the predictions of
stellar models by altering the mass-loss rates assumed in the stellar
models. By comparing to observations we can constrain which rates best
fit the observation.  Similarly, varying the rotation rate does change
the evolution of stars by extending their main-sequence
lifetimes. Also the stars tend to have higher surface temperatures as
they evolve. Therefore there are changes to the shape of the
populations on the HR diagram as they age. \cite{cyclist}.

\subsection{Binary stars}

\begin{figure}[t]
\sidecaption[t]
\includegraphics[scale=.5]{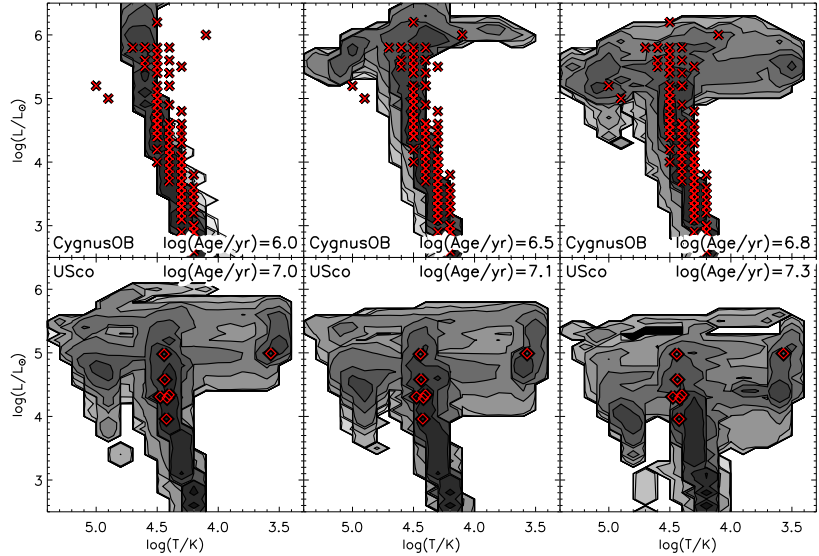}
\caption{Similar to Figure \ref{fig3} but now with a synthetic binary
  star population compared to the observed populations of two star
  clusters, the Cygnus OB association\cite{wright} in the upper panels
  and USco association \cite{pecaut} in the lower panels. The age of
  the population can be estimated by comparing the distribution of the
  most luminous stars but is more uncertain in the case of binary
  populations. Each contour is separated by an order of magnitude in
  the probability density of stars on the HR diagram.}
\label{fig4}     
\end{figure}

All the details that are required to make a single-star synthetic
population also apply for a binary population. However now we need to
worry about more than just the IMF. There
must also be an initial distribution for how many binaries there are,
as well as the orbit and the mass of the second star. The simplest
assumption is that there is a distribution in the mass ratio of the
secondary to the primary star. This is normally defined as $q=M_2/M_1$
and is assumed to range from 0 to 1 with a distribution of
$f(q)=q^{\gamma}$, typically with $\gamma=0$. The second detail
required is orbital separation or period distribution. The simplest
assumption here is \"Opik's law which assumes a flat distribution in
$\log P$. Yet the number of binaries in the population has no real
theoretical basis and must be measured.

For many years it has been difficult to measure these parameters. The
current understanding is that for massive stars above 8$M_{\odot}$
60\% are in multiple systems, while above 16$M_{\odot}$ this increases
to 80\%. Furthermore a small number of these multiples include more
than one companion star. In comparison only a quarter of the lowest
mass stars are in multiple systems \cite{duchenekraus2013}. For most
mass ranges $\gamma$ is consistent with being flat. For massive stars
$\gamma=-0.1 \pm 0.6$ although for systems wider than 100AU we find
that $\gamma=-0.5 \pm0.1$ indicating a preference for more unequal
mass ratios. The period distribution is also consistent with \"Opik's
law although there is also a slight excess of systems with 5 day
periods in this distribution \cite{duchenekraus2013}. Other mass
ranges tend to have typical orbital separations with smaller orbits
for lower mass stars.

Finally the eccentricity distribution is slightly biased towards
circular orbits with a range of eccentricities up to close to
1. Approximately circular systems are more likely and a recent
estimate is that, $dN/de \approx 0.4+1.2e$
\cite{tokovininkiyaeva2015}. It is uncertain how important the
eccentricity is for the evolution of binary systems. It has been
suggested that tides circularise the orbits before interactions
\cite{hurley}. Thus the semi-latus rectum distribution is the most
important initial parameter. Conversely in some cases eccentricity
during mass transfer may be important for determining the outcome of
the mass transfer \cite{kalogera1}. It is clear, that closer
binaries are less likely to be eccentric than the wider binaries that
represent the effect of tides circularising the binary.

In light of knowing these distributions we are now left with the
question of how to sample this parameter space with stellar evolution
models. There are the two widely used techniques of creating binary
stellar models, rapid and detailed models. For the rapid method a full
sampling of the initial parameter space is possible. The initial
masses, periods and eccentricities can be sampled at random or
uniformly from the input distributions. The rapid models can be
calculated in a fraction of a second, thus many models could be
calculated to simulate a population. These models necessarily make
approximations and apply analytic prescriptions for some aspects of
stellar evolution \cite{han,hurley,scenariomachine}.

Only recently it is possible to calculate large numbers of detailed
models, where the full stellar structure is computed, due to the
increase of computational resources. A detailed model grid was first
calculated by the Brussels group \cite{vanbev}. Now more groups are
also taking up this method \cite{bpass,marchant} gaining accuracy in
the models of stellar structure and evolution. However because models
take minutes to run the computational cost is a few orders of
magnitudes greater than the rapid models. Thus their ability to search the
parameter space is severely constrained. It is therefore important for
the detailed models to sample the space over a uniform grid.

One further detail must be accounted for during the evolution of the
binary system: what happens after the first (and second!)
supernova. The neutron star kick can destroy the binary or keep it
bound. The kicks are modelled as discussed above but the full range of
directions and kick strengths must be sampled. While the kicks are
uncertain, increasing accuracy in supernova simulations and
observations provide growing evidence that neutron star kicks may be
less random than currently suggested \cite{braykicks,snsims}. Black
hole kicks are a little more uncertain, mainly due to the lack of
observations of single black holes. Many seem to suggest that there
are at least weak kicks \cite{bhkickslow}.

Creating detailed models after the first supernova is tricky due to
the large number of possible outcomes for each individual model. This
is when the rapid evolution models have the advantage because it is
possible to sample all the probable future evolutionary pathways in
the same time it takes to blink. It is also possible to investigate
the effect of all the uncertainties, such as initial parameters,
common-envelope evolution model and supernova kick distributions. This
provides an understanding of where we should focus our efforts on
reducing those uncertainties.

Once the models are calculated, again they must be processed to
predict the details of the synthetic population for comparison to
observations. A similar method can be used as before and we plot in
Figure \ref{fig4} a new version of Figure \ref{fig3} but for a binary
population. We see the predictions change and fill a larger parameter
space. There is no longer a single isochrone for a single-aged
population. The processes of RLOF and CEE cause some of the stars to
evolve on different paths. Importantly we see that the derived ages
become slightly different for the binary population. In general binary
populations appear bluer and more luminous at the same age than
single-star populations. This suggests at least for Cygnus OB the age
could be up to 6Myrs rather than the 3Myrs suggested by single star
models.

The key evidence that binaries are important for population synthesis
come from the direct observational evidence from X-ray binaries,
double neutron-star binaries and gravitational waves from black-hole
mergers. The observation of such systems indicates that many
supernovae must also happen within binary systems and there is growing
observational evidence of this from supernovae themselves.

\section{Binary population synthesis of supernovae and their progenitors}

In this section we highlight the inferences that are important for
understanding supernovae from population synthesis.

\subsection{Type Ia thermonuclear supernovae}

From the early stages of supernova studies it was realised that some
of the events with no hydrogen arose from a different stellar
population \cite{tinsley}. It was an old rather than young stellar
population indicating that the progenitors must have been a few 100
Myrs or older. Eventually it was realised that the explosion of a
white dwarf at the Chandrasekhar limit could explain the energy
released in the supernovae and explain the uniformity of the
explosions.

A white dwarf is the burnt out remnant of an intermediate mass star
within which no further nuclear fusion reactions are possible. The
objects are supported by electron-degeneracy pressure. It was
Subrahmanyan Chandrasekhar who first combined special relativity and
quantum mechanics to show that there was a maximum mass for a white
dwarf. When the mass reaches or exceeds this limit the star cannot be
supported and collapses. In the case of a carbon-oxygen white dwarf
this can lead to the ignition of explosive nuclear fusion of carbon
followed by a thermonuclear supernova \cite{whelan,iben,webbink}.

We mention these as they are an important result of binary evolution
and may provide firm constraints in the future. There is also some
overlap between what might be type Ia and core-collapse events.  For
example the type Iax events have been suggested to be either
core-collapse or thermonuclear events. For these events a possible
progenitor has been detected. Supernova 2012Z had its progenitor
observed as a faint blue star \cite{2012Z} suggesting the
progenitors are possibly helium stars transferring matter onto an
exploding white dwarf.

\subsection{Core-collapse supernovae}

\begin{figure}[t]
\sidecaption[t]
\includegraphics[scale=.39]{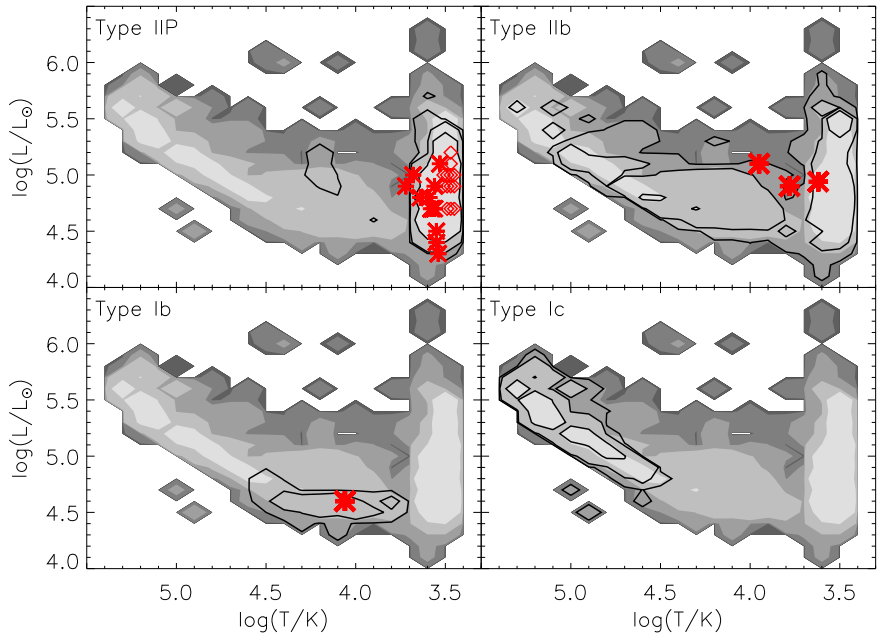}
\caption{Example Hertzsprung-Russell diagrams with the predicted
  location of supernova progenitors compared to observed
  progenitors\cite{smartt2015}. Grey contours are for all progenitors
  while the black contours represent the location of a specific
  supernova type as labelled in each panel. The asterisks in each
  panel represent the location of observed progenitors and the
  diamonds are upper limits.}
\label{fig5}     
\end{figure}

The more common type of supernovae arises from the final core-collapse
of massive stars. These are stars that are initially more massive than
8$M_{\odot}$ and in which nuclear fusion progresses all the way to
form an iron-group element core. Further nuclear fusion releases no
energy so the core collapses to a neutron star or black hole depending
on the mass of the core. This collapses releases a tremendous amount
of energy that causes the star to explode. While this general picture
holds, some of the exact details are still being worked out. Only
recently supernova simulations have been able to lead to model
explosions.

The core-collapse supernovae fall into two principal types, type II
and type Ib/c which are hydrogen-rich and hydrogen-free
respectively. Population synthesis' key role is in predicting the
relative populations of these two supernovae as they depend primarily
on the amount of mass loss the progenitor stars experiences before
they explode. For single stars this is dependent on their stellar
winds, while for binary stars it is the result of CEE and RLOF.

There is a strong consensus from multiple authors that interacting
binaries must dominate the rate of observed core-collapse supernovae
\cite{bpass,sne2,sne3,sne4,sne5,e13}. All these studies show that
binaries increase the number of hydrogen-free progenitors at the
expense of red supergiant, hydrogen-rich progenitors. This is a common
feature of nearly all binary population predictions concerning
core-collapse supernovae.

One important feature that population synthesis can predict is how the
observed ratio of supernova types varies with metallicity. It was
initially thought that the relative SN rates should be constant with
binary interactions but vary for single stars because stellar winds
are radiatively driven and thus metallicity dependent. The fact is
that binary interactions only remove some, not all, of the hydrogen on
the surface of a star and the rest must be removed by stellar
winds. This leads to more type II SN relative to type Ib/c SNe at
lower metallicities.

While the rates provide an important constraint on binary stellar
population synthesis, a more direct indication is the growing number
of observed progenitors of core-collapse SNe. The next step for
population synthesis is to model the distribution of progenitors in
the HR diagram as well as the relative rates of the different types of
SNe. We show in Figure \ref{fig5} the predicted location of the
different core-collapse SNe compared to the locations of observed SN
progenitors. The most commonly detected progenitors are the type II
events while only one type Ib supernova progenitor is detected. The
detected progenitors also show surface temperatures of below 10,000~K;
likely because hotter stars output most of their flux in the
ultraviolet and are also more affected by dust which make them
difficult to detect in pre-explosion images \cite{e13,yoon}.

Studying the progenitors has revealed that the type II supernovae lack
the most luminous and massive predicted progenitors. In comparison the
progenitors of type Ib/c supernovae all remained unobserved despite
the expected progenitors, Wolf-Rayet stars, being some of the most
luminous. One problem is that while they might have a high bolometric
luminosity most light is emitted in the ultraviolet so they are also
faint in the optical. A population synthesis model showed that the
non-detections point towards most type Ib/c actually came from binary
progenitors. It was also suggested \cite{yoon} that some of the
progenitors from binaries, being cooler, should also be easier to
observe. This was borne out by the detection of the first type Ib
progenitor that (after a lot of competing studies) was shown to be
most likely a helium giant evolved from a lower mass star
\cite{cao,bersten,groh,e16}.

This leaves us with two unresolved questions, where are these
predicted helium stars in our Galaxy and what happens to the
Wolf-Rayet stars? Some of the Wolf-Rayet stars may still explode but
some may simply collapse directly to a black hole and have no
supernova, as some of the more massive red supergiants have been
suggested to \cite{smartt2015}. Population synthesis cannot be used to
resolve these questions yet as they depend too much on the
uncertainties of CEE, RLOF and core-collapse itself.

One pathway that may be able to provide some answer to this question
in future. A consistent result between binary population synthesis
codes is that binary interactions extend the delay-time range for
supernovae \cite{zapartas}. The delay time is the time between a star
being formed and exploding in a supernova. In a single star population
the range would be from the lifetime of a 100$M_{\odot}$ star to an
$\approx8M_{\odot}$ star. Or between 3Myrs to 41Myrs. However for a
binary population the interactions can lead to core-collapse supernova
at ages up to 200Myrs after the stars were formed. This may not be a
small effect with predictions of between 8.5\% to 23\% of supernova
occurring after the single-star cut-off age. The exact number of
``late'' supernovae depends on the uncertainties in the population
synthesis but all the codes to predict them. Future study of delay
times will give us further evidence of the importance of binary
interactions in supernovae but also possibly answer more about the
nature of the progenitors for all supernova types.

Finally other observations that can be modelled and place constraints
on synthetic populations involve modelling the outcomes of a binary in
a supernovae. Much work for example has been done to model X-ray
binaries \cite{bhkicks,bhkickslow,belczynski}, runaway stars and also
long-gamma ray bursts that are associated with supernovae
\cite{cantiello,sne4,grb1}. Combining multiple observations in
future will provide greater insight into CEE and especially the
creation of kicks for forming neutron stars and black holes.

\subsection{Compact remnant mergers}

Once the evolution driven by nuclear fusion is over for the stars in a
binary the evolution becomes relatively simple. The stellar remnants,
be they white dwarfs, neutron stars or black holes, are (to a good
approximation) unchanging. This is somewhat untrue for white dwarfs as
they do continue to cool, as do neutron stars to a smaller degree. The
only remaining event to occur for such binaries is an eventual merger
due to the tightening of the orbit by emission of gravitational
radiation. The history of predicting mergers of compact remnant
binaries is long.

Gravitational radiation is a natural consequence of the General Theory
of Relativity and how the orbit and eccentricity of a binary were
affected by gravitational waves were derived by Peters \cite{peters1964},
\begin{equation}
\frac{da}{dt}=-\frac{64 G^3 M_1 M_2 (M_1+M_2)}{5 c^5 a^3 (1-e^2)^{7/2}} \left(1 + \frac{73}{24}e^2 + \frac{37}{96} e^4 \right)
\end{equation}
and
\begin{equation}
\frac{de}{dt}=-\frac{304 e G^3  M_1 M_2 (M_1+M_2)}{15  c^5 a^4 (1-e^2)^{5/2}} \left(1 + \frac{121}{304}e^2 \right)
\end{equation}

In the case of a circular orbit the time taken for a merger is given by,
\begin{equation}
T_{\rm m,e=0}(a)=\frac{5 c^5a^4 }{256 G^3 M_1 M_2 (M_1+M_2)}
\end{equation}
while for a highly eccentric binary with $e$ close to one,
\begin{equation}
T_{\rm m,e\sim1}(a,e)=\frac{768}{425} T_{\rm m,e=0} (1-e^2)^{7/2}
\end{equation}

Using these equations it is straightforward to calculate how long it
will take two compact remnants to merge. In many cases, due to the
weakness of the gravitational radiation, the time to merge will be
greater than the age of the Universe and so merger rates are relatively
low, although interactions with a third body may increase the chance
of a merger.

Before 2015 the only events that were observed and likely to be the
result of such mergers were type Ia supernovae and short-GRBs, at
least some of which are thought to be the result of white dwarf or
neutron star mergers respectively. On the 14th of September 2015 the
merger of two massive black holes was detected via the gravitational
waves released during their final inspiral and merger
\cite{gw150914}. This gave the final direct evidence that such mergers
do occur. The fact that the first detection would be a binary black-hole
merger was predicted by \cite{belczynski1} and is no surprise
considering that more massive black holes have a shorter merger time
and also a stronger Chirp mass ($\mathcal{M}_{0}=(M_1
M_2)^{3/5}/(M_1+M_2)^{1/5}$). This makes them more likely to be
detected as the the detection distance is $\propto \mathcal{M}_{0}
^{5/6}$.

While the time it will take two compact objects to merge is
straightforward to calculate given their masses and initial period,
the prediction of the initial parameters of the compact remnant
binaries is extremely uncertain. To predict it a stellar population
must be modelled from birth to the final death-throes of both
stars. Therefore \textit{every} uncertainty in stellar evolution and
population synthesis has an effect on the final merger population, as
has any uncertainty in the star-formation history of the
population. If the uncertainties of evolution are varied over a
reasonable range, the final predicted results can span a large range
\cite{belczynskianddemink} even though most models predict similar
orders of merger rates
\cite{bhbh3,bhbh4,bhbh2,marchant,bhbh5}. Modelling the rate we would
observe today from all compact remnant binaries formed in the Universe
requires understanding of the mass and star-formation history of the
Universe and how the metallicity of the Universe varies with the
cosmic age \cite{bhbh2}.

The key problems are making enough close binaries with compact
remnants and making enough massive black holes. For the remnants to
merge within the age of the Universe they have to be massive, with
their orbits highly eccentric and/or very close. In future as more
merging compact remnants are detected the sensitivity of the models
will be useful. As the population of observed mergers becomes
understood there will be tighter constraints on the possible range of
the input population synthesis parameters as well as possibly the
star-formation history of the Universe. In addition it can be hoped
that gravitational waves will be detected from neutron star mergers
that may also be associated with short-GRBs or other bright optical
transient sources. This will reveal not only a great deal about neutron star
structure but also about the uncertain kicks in their natal
supernovae.

\section{Conclusions}

Binary population synthesis has achieved a great number of success in
predicting details of observed populations but is still highly
dependent on a number of uncertainties in both binary stellar
evolution and the construction of synthetic stellar
populations. However in light of the very final merging of two compact
remnants associated with binary stars detectable by gravitational
waves, we may be approaching a time when unique and strong constraints
can be placed on the uncertain physics.

\begin{acknowledgement}
The author would like to thank J.C.C. Wang, E.R. Stanway, L. Xiao,
L.A.S. McClelland and J.C. Bray for proofreading and commenting on
this chapter.
\end{acknowledgement}

\section*{Cross-References}
\begin{itemize}
\item Gal-Yam, A. \textit{Observation and Physical Classificiation of Supernovae}
\item Arcavi, I. \textit{Hydrogen-Rich Core Collapse Supernovae''}
\item Pian, E., Mazzali, P. \textit{Hydrogen-Poor Core Collapse Supernova}
\item Limongi, M., \textit{Supernovae from massive stars (12-100 Msun)}
\item Benvenuto, O. Bersten, M., \textit{Close Binary Stellar Evolution and Supernovae}
\item Van Dyk, S., \textit{Supernova Progenitors Observed with HST}
\item Casares, J., Israelan, G., Jonker, P., \textit{X-ray Binaries}
\item van der Heuvel, E., \textit{Supernovae and the Evolution of Close Binary Systems}
\end{itemize}

\end{document}